# Mössbauer studies of powdered single crystals of FeTe$_{0.5}$Se$_{0.5}$

K. Szymański[1], W. Olszewski[1], L. Dobrzyński[1,2], D. Satuła[1], D. J. Gawryluk[3],

M. Berkowski[3], R. Puźniak[3], A. Wiśniewski[3]

[1]University of Bialystok, Faculty of Physics, Lipowa 41, 15-424 Białystok, Poland

[2]The Andrzej Sołtan Institute for Nuclear Studies, 05-400 Otwock-Świerk, Poland

[3]Institute of Physics, Polish Academy of Sciences, Al. Lotników 32/46, 02-668 Warsaw, Poland

E-mail: satula@alpha.uwb.edu.pl

**Abstract.** The Mössbauer measurements performed on powdered single crystals of FeTe$_{0.5}$Se$_{0.5}$ ($T_c \sim 14.7$ K) reveal minor content of two impurity phases, identified as Fe$_3$O$_4$ and Fe$_7$Se$_8$, among the major tetragonal phase. From the shape of impurity subspectra it follows that Fe$_7$Se$_8$ behaves in superparamagnetic manner, most likely because of randomly distributed Fe vacancies in the lattice structure of Fe-Te-Se. The magnetite content in the powdered absorber exposed to ambient air conditions remains unchanged during period of 16 months. Ageing effects were observed for the samples stored under argon atmosphere and small increase of the isomer shift of the doublet was detected. Presented temperature dependence of the hyperfine parameters can be explained as due to possible orthorhombic distortion or to temperature behaviour of impurity phases Fe$_3$O$_4$ and Fe$_7$Se$_8$. Strong tendency to formation of crystalline texture of powdered sample is observed.

## 1. Introduction.

Very soon after the discovery of superconductivity in LaFeAsO$_{1-x}$F$_x$ pnictides with $T_c \sim 26$ K [1], superconductivity was found in chalcogen-iron system FeSe$_{1-\delta}$ ($T_c \sim 8.5$ K) [2] with the simple chemical formula and the same layered tetrahedral crystalline motif. McQueen *et al.* [3] reported very narrow range of $\delta$ value, $\delta = 0.01 - 0.03$, in which the material shows the superconductivity. Williams *et al.* [4] confirmed that only FeSe$_{0.99}$ contains the tetragonal phase without iron precipitations as well as without presence of additional Fe$_7$Se$_8$ phase. These reports indicate extreme sensitivity of superconductivity in Fe-Se system to the appearance of any defects in the crystal structure and of any impurities, such as Fe$_3$O$_4$ or Fe$_7$Se$_8$.

Various chemical substitutions to Fe-Se system were done. It was found that substitution of a part of selenium by tellurium leads to isostructural, pseudobinary system of Fe-Te-Se with higher $T_c$ up to ~ 15 K [5, 6]. In many reports on Fe-Te-Se system, the authors noticed magnetic anomaly at about 120 K [7, 8], seen in electrical [9] or thermal [8, 10] transport measurements. Very often the anomaly was linked to the Verwey transition [11] in Fe$_3$O$_4$ [8]. Moreover, impurities of Fe$_7$Se$_8$ were detected in structural [12] or the Mössbauer measurements [7]. However, it was concluded that the impurities whose content does not extend beyond 3% do not affect superconducting properties of Fe-Te-Se system, contrary to the impact of such phases on the superconductivity in Fe-Se system. On the other hand, in the reports describing substitution of Fe by Co, Ni and Cu ions, strong suppression of superconductivity in both Fe-Se and Fe-Te-Se systems was noticed even with a substitution as small as 1 - 2% [13, 14].

In order to better understand the local structure of Fe-Te-Se system and the role of the impurities and defects in the formation of the superconducting state we performed Mössbauer studies of powdered single crystals of $FeTe_{0.5}Se_{0.5}$. The Mössbauer technique is sensitive to the hyperfine magnetic field sensed by $^{57}Fe$ probe and thus coexistence of a magnetic ordering and superconductivity could be investigated. This type of studies demonstrates suppression of magnetic ordering by substitution of a part of oxygen by fluorine in LaFeAsO [15] and a spin wave anomaly in $BaFe_2As_2$ [16]. On the other hand measurements of the recoilless fraction reveal anomalies in iron dynamics recently observed in LiFeAs [17] and $FeTe_{0.5}Se_{0.5}$ [7].

## 2. Experimental details

Superconducting single crystals of $FeTe_{0.5}Se_{0.5}$ have been grown using the Bridgman's method. The samples were prepared from stoichiometric quantities of iron chips (3N5), tellurium powder (4N) and selenium powder (pure). Double-walled evacuated sealed quartz ampoule containing the starting materials was placed in a furnace with a vertical gradient of temperature. The samples were synthesized for 6 h at 680°C, then the temperature was increased up to 920°C. After melting, the temperature was held for 3 h, then the samples were cooled down to 400°C at the rate of 3°C/h and next to 200°C at the rate of 60°C/h, and finally cooled down to room temperature with the furnace. The obtained crystals exhibit a natural cleavage plane (001). They were superconducting and exhibited sharp transition to the superconducting state with critical temperature about $T_c \sim 14.7$ K. The details of superconducting properties, structural and chemical analysis of the grown crystals have been published elsewhere [18].

Bulk pieces of $FeTe_{0.5}Se_{0.5}$ compound were manually grounded to a powder under argon atmosphere in a glove box. We have checked that grinding (at room temperature) does not affect $T_c$. Absorbers for the Mössbauer measurements were prepared by distributing the powder on a scotch tape. Planar mass density of the powder was measured. A standard spectrometer working in the constant acceleration mode, equipped with a closed-cycle refrigerator, was used for the Mössbauer measurements. Lowest temperature available in our experimental set-up was 13.3 K. During these measurements the absorbers were kept inside an evacuated cryostat in order to protect the samples from oxidation. In order to determine whether the samples exhibit any oxidation effect, some of the absorbers were exposed to air and measured systematically at various times since exposition, see figure 1. Clear-cut evidence of the presence of small amount of magnetite follows from the observed spectra when one compares the spectrum of 16 months old sample with the spectrum of a sample with intentionally added minute quantity of magnetite (figures 1 (e) and 1 (f)). In order to observe ageing of samples, remaining part of freshly prepared powder was stored under argon atmosphere and after 13 months new absorbers were prepared for the second series of the Mössbauer experiments. This part of the experiment has been carried out using a new source with larger activity.

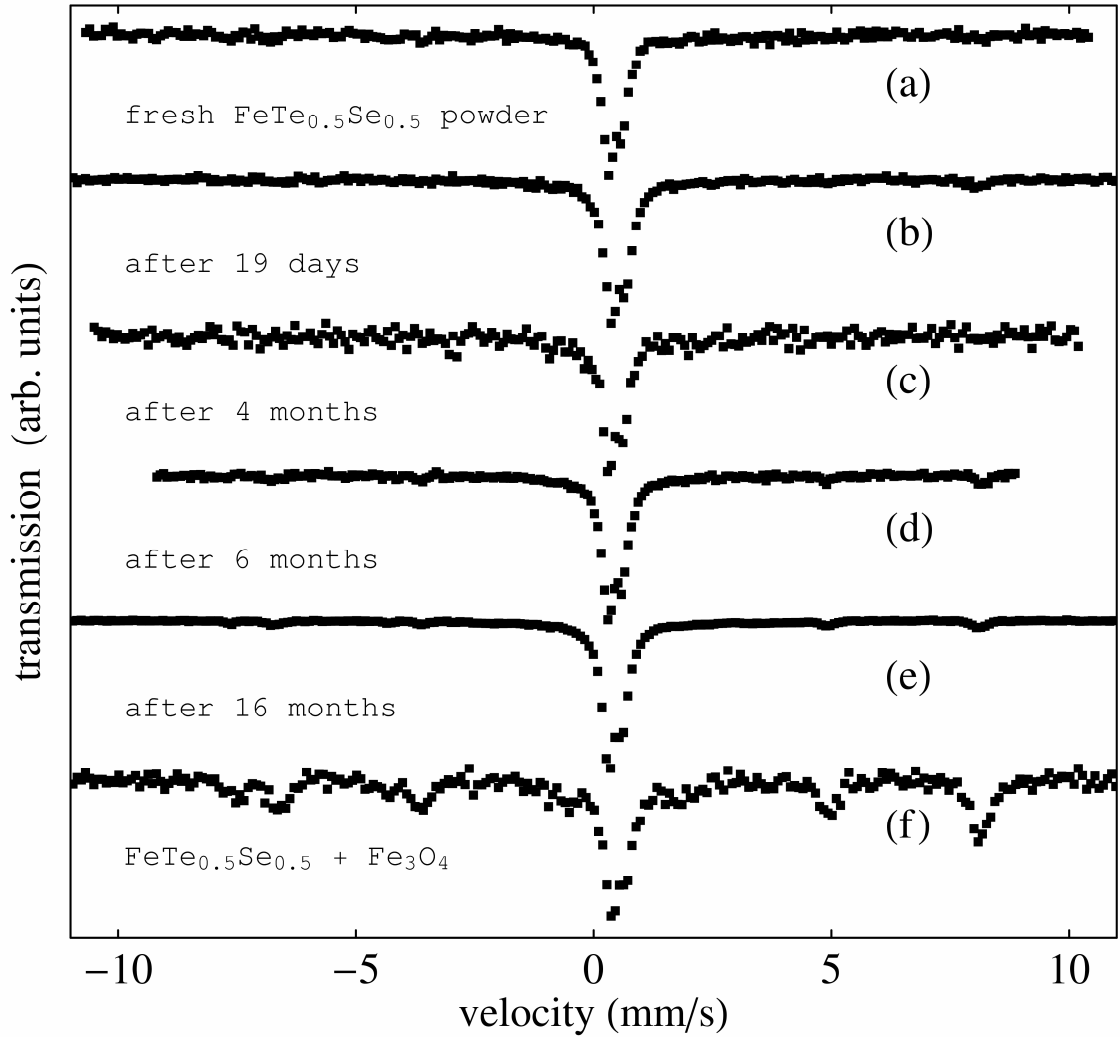

Figure 1. The Mössbauer spectra of the FeTe$_{0.5}$Se$_{0.5}$ with planar density 21(1) mg·cm$^{-2}$: (a) for freshly prepared absorber and for the absorber exposed to air for (b) 19 days, (c) 4 months, (d) 6 months, (e) 16 months. (f) The same absorber measured together with extra 3.6 mg·cm$^{-2}$ Fe$_3$O$_4$.

In order to demonstrate to which extent mechanical grinding introduces additional phases to the parent compound and therefore influences the content of minority phases in the studied materials, magnetization measurements were performed at temperature above transition temperature to the superconducting state on a piece of single crystal, on fresh powder obtained by grinding of its part under argon atmosphere and on powder obtained by grinding of its part in air at ambient conditions. The measurements, using Princeton Applied Research (model 4500) vibrating sample magnetometer (VSM), were completed at the temperature of 20 K and in magnetic fields up to 6 kOe. The results are shown in figure 2. All powdered samples were obtained in hand mortar. It is not strictly recurrent process. We did not control the grain size in the grinding process and therefore we can not exclude that the different hysteresis loops may stem from different grain sizes.

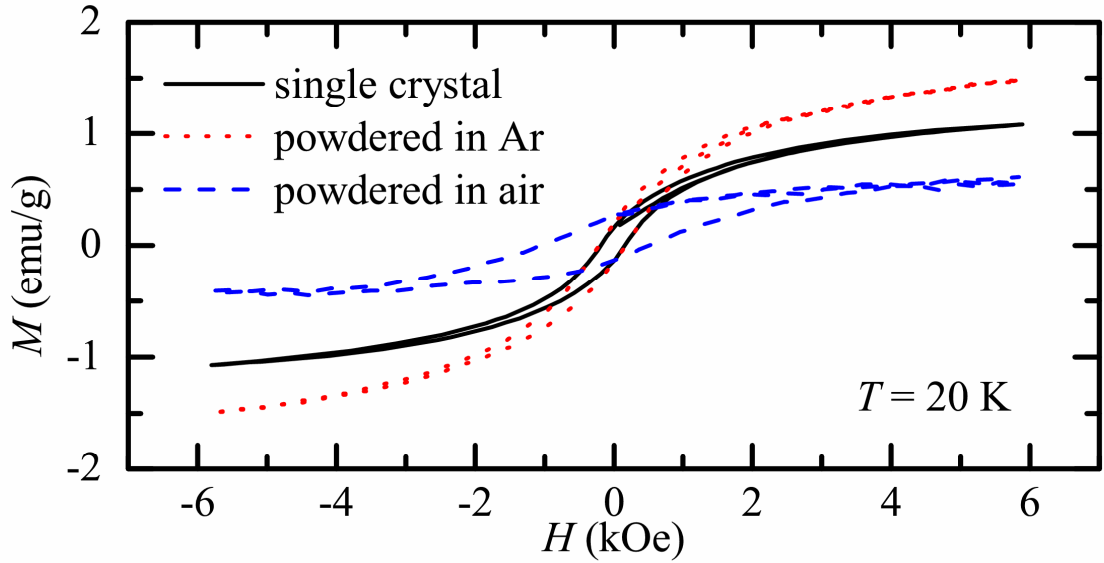

Figure 2. Influence of grinding on the content of minority magnetic phases in the studied materials. Magnetization hysteresis loops of FeTe$_{0.5}$Se$_{0.5}$ single crystal and powder obtained by grinding under argon and in air at ambient conditions. The data were recorded at 20 K in magnetic fields of −6 to +6 kOe.

**3. Experimental results**

The Mössbauer spectra of the absorbers exhibit shape of an asymmetric doublet. The powdering of the samples resulted in powder grains with flake-like shapes that could be hardly changed. Therefore, the samples, glued to a scotch tape, exhibited inherent texture. Its value is remarkable indeed. How much the relative ratio of the line intensities is sensitive to the sample orientation with respect to the gamma wave vector direction is demonstrated in figure 3. This offers a possibility of carrying out the data analysis similarly to that one used in the case of a single crystal data, and to obtain in this way additional information, unavailable for powder samples without preferred orientations. We have also performed so called "magic angle" [19] Mössbauer measurements to observe spectrum of a texture free sample, see figure 3 (c). The spectrum is symmetric doublet and we could not detect any additional component reported in [7].

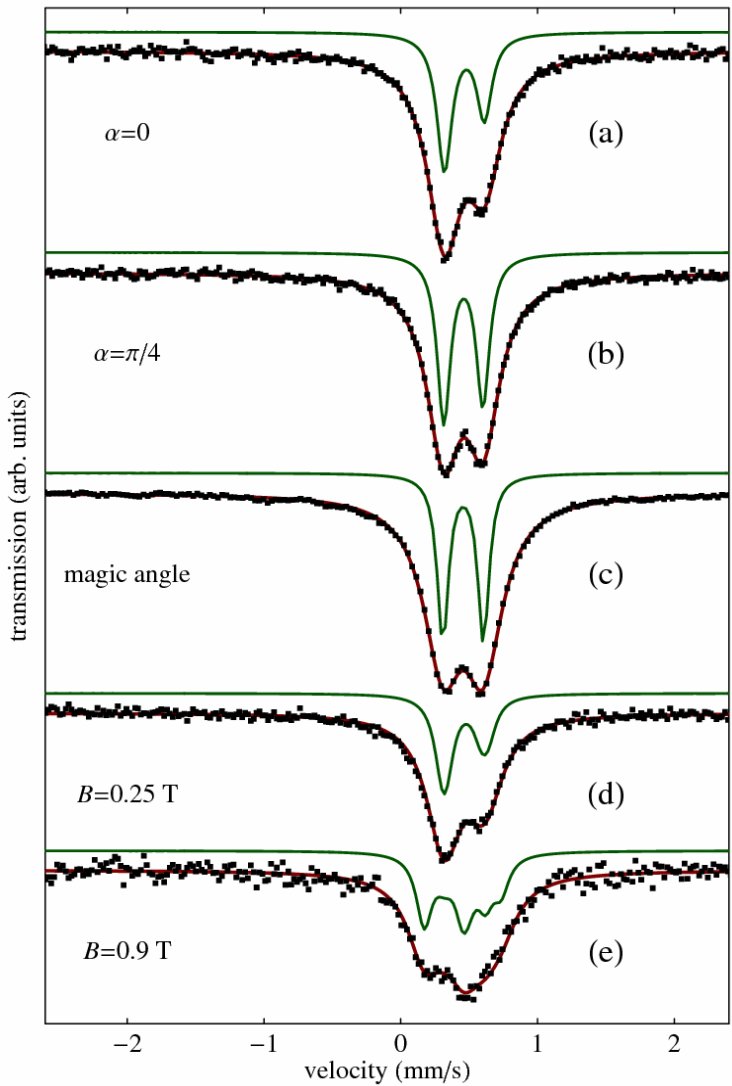

Figure 3. The Mössbauer spectrum measured with wave vector of radiation (a) normal to the sample plane (b) inclined by π/4 from the normal direction, (c) in "magic angle" arrangement. Absorber measured in external magnetic field, induction $B$ applied (d) parallel and (e) perpendicular to the sample plane (wave vector of radiation was normal to the sample plane in (d) and (e)). Planar density of the mass was equal to 21(1) mg·cm$^{-2}$. The components shown above experimental points represent the resonant cross section with the line width reflecting the absorber line width only.

Using the transmission integral [20] in which the source width was determined in separate calibration procedure, all measured spectra, with different sample orientation, and with and without applied magnetic field, were fitted with single component - the doublet. Full Hamiltonian treatment [21] was applied to the data. The quality of such fits can be seen in figure 3.

In the fitting procedure the asymmetry parameter of the electric field gradient (EFG) was set to zero because the local symmetry of Fe atoms in the layered structure can be safely assumed to be axial. Only two parameters, namely, the main component of the EFG and the angle between wave vector and the $z$-axis of the EFG were fitted. In the case of single crystal with local axial symmetry, this angle should be zero. In our case, the angle turned out to be in the range between 30 and 45 degrees, directly related to the crystalline texture. Smaller intensity of the absorption line located at higher velocity indicates the negative sign of the EFG. The value of the nuclear quadrupole moment of the excited, $I = 3/2$, state of $^{57}$Fe nucleus was taken from [22]. Because our data were approximated by functions related to a single crystal, we have performed additional measurements in external magnetic field, see figures 3 (d), (e). Again we were able to fit the single component only. The values of magnetic induction agree within 10% with the values measured by the Hall probe, and the sign of the quadrupole splitting is negative, consistently with the results obtained in zero magnetic field.

Temperature measurements down to 13.3 K reveal neither magnetic splitting nor line broadening of the doublet. By fitting the spectra with one asymmetrical doublet it was possible to obtain temperature dependence of quadrupole splitting, centre shift, full width at half maximum of the Lorentzian lines and the ratio of areas under the lines forming asymmetrical doublet shown in figure 4. This has been done for the spectra obtained in experiments performed on fresh powder and aged powder stored under Ar atmosphere as well. The data for these two experiments show small yet systematic deviations. The most important point is that quadrupole splitting (figure 4(a)) measured in the experiment with higher statistics on aged sample shows abnormalities and they correspond to the temperatures at which abnormalities were observed in magnetic susceptibility measurements of the $FeTe_{0.5}Se_{0.5}$ samples (figure 5), magnetite [23] and $Fe_7Se_8$ [24]. The origin of the peak in the imaginary part of susceptibility visible in figure 5 at 50 K is unclear. However, most likely it could be linked with spin density anomaly. Such anomaly was observed in transport measurements [5]. In the undoped, non-superconducting, parent Fe-Te system the long-range spin density wave (SDW) order was observed near 65 K [12, 25, 26]. The short-range magnetic order in non-superconducting single crystals of $Fe_{1+y}Te_{1-x}Se_x$ ($x$ = 0.1 - 0.3) [27] as well as the short-range magnetic order (weaker with increasing Se concentration) in superconducting single crystals with $x$ = 0.25 and 0.3 was observed too [28]. Small systematic deviation of the centre shift of the fresh and aged samples (figure 4(b)) is observed. Data with sufficiently high statistics taken within broad range of velocities were accumulated to observe quantitatively amount of the impurity phases, see figure 6. An asymmetric doublet and subspectra related to $Fe_7Se_8$, $Fe_3O_4$ and $\alpha$-Fe possible phases were fitted to the spectra. Because amount of the impurity phases is small, their hyperfine parameters were fixed. In particular, the values given in [24] and [29] were used to construct subspectra of $Fe_7Se_8$ and $Fe_3O_4$, respectively. Initial fits show that content of $\alpha$-Fe is below 1%. Thus in further analysis the content of $\alpha$-Fe was assumed to be zero; results of the best fits are shown in figure 6 and quantitative results are collected in table 1.

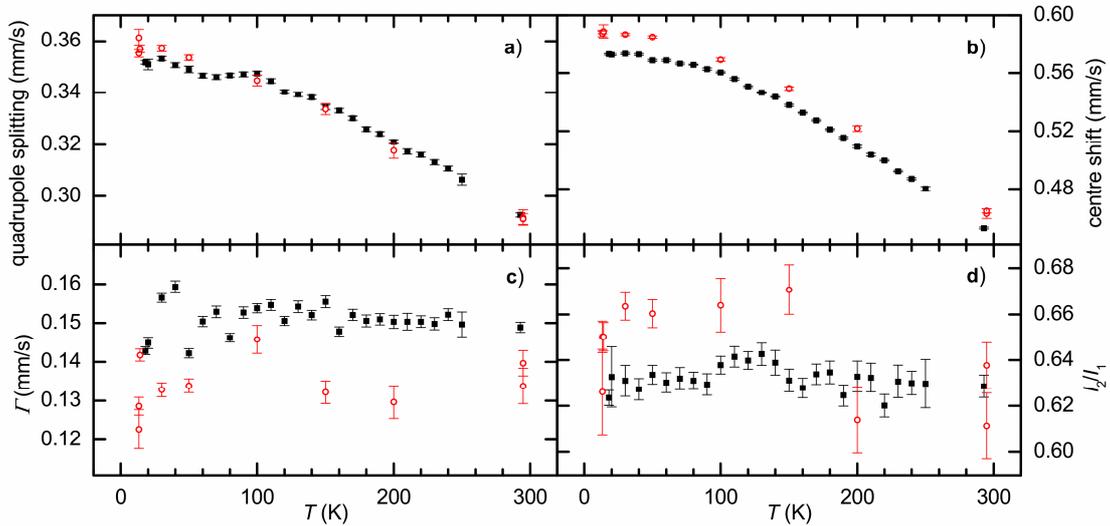

Figure 4. Temperature dependence of the parameters of main doublet. Open symbols correspond to measurements on freshly prepared powder with planar density of the mass equal to 21(1) mg·cm$^{-2}$ while full points show results obtained for aged sample in Ar atmosphere (13 months) with planar density of the mass equal to 23(1) mg·cm$^{-2}$.

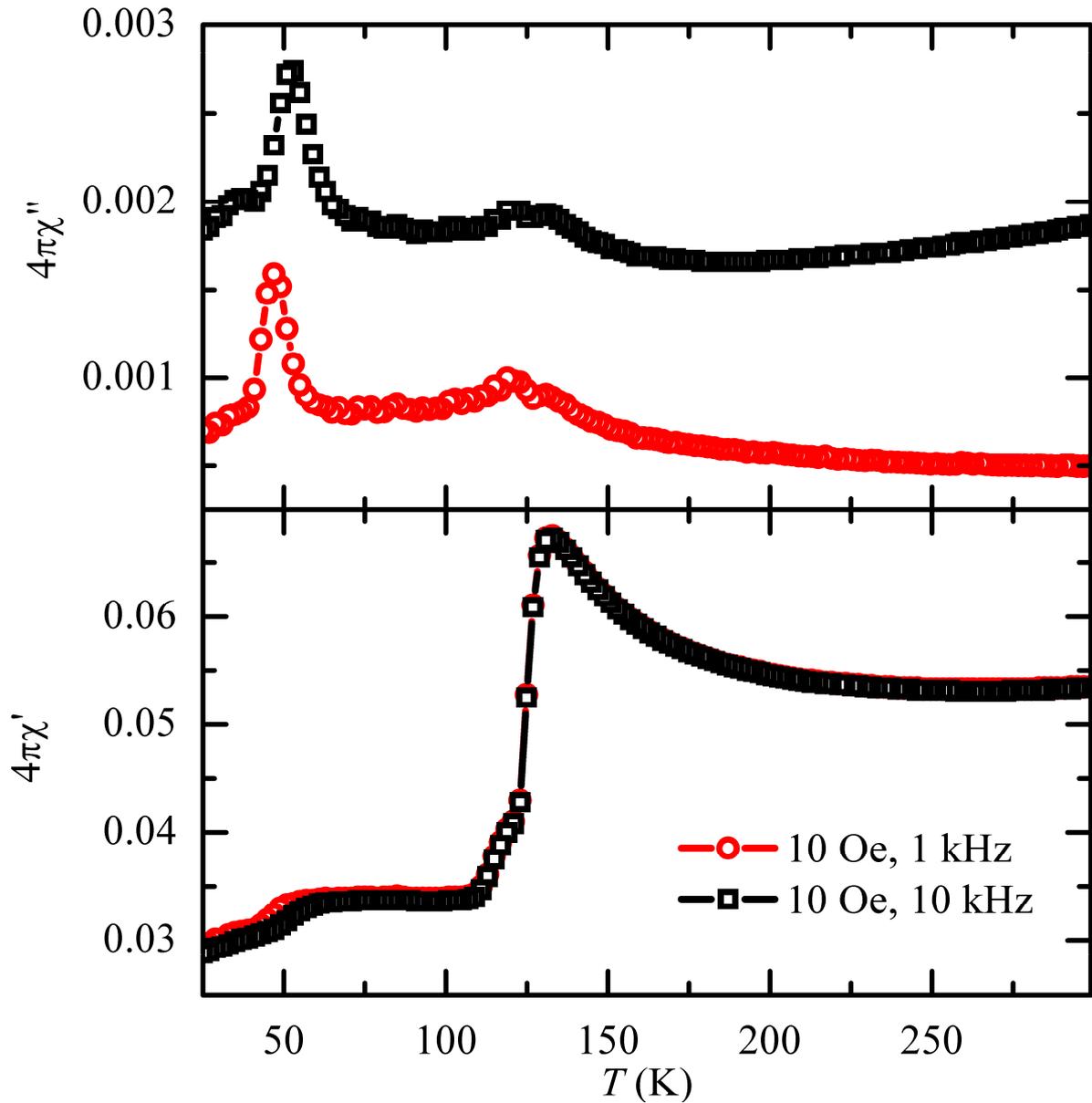

Figure 5. Temperature dependence of real part (4πχ' - lower panel) and imaginary part (4πχ'' - upper panel) of AC susceptibility for $FeTe_{0.5}Se_{0.5}$ single crystals, measured above superconducting transition temperature in 10 Oe of AC field with 1 and 10 kHz in warming mode.

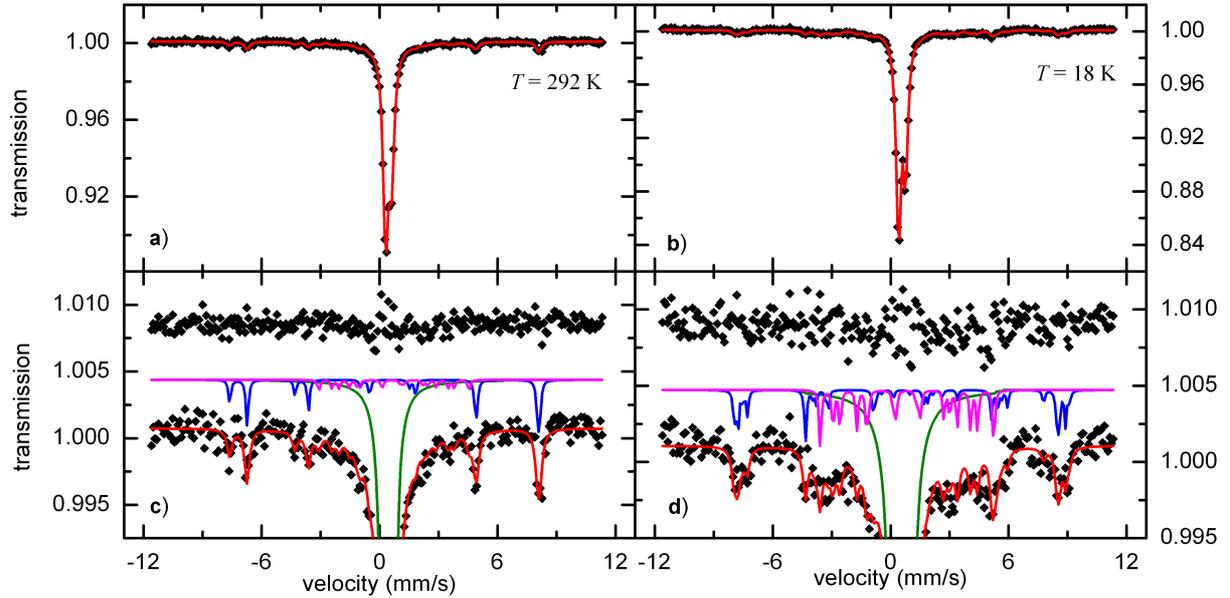

Figure 6. The Mössbauer spectra of FeTe$_{0.5}$Se$_{0.5}$ samples aged in Ar, measured at (a) room temperature and (b) $T = 18$ K temperature, fitted with four components. In (c), (d) the spectra are shown with vertical scale enlarged and the four components are shown separately: asymmetric doublet (most intense component), Fe$_7$Se$_8$ (moderately split component), Fe$_3$O$_4$ (mostly split component). The central data sets are difference plots between actual spectra and fit curves.

Table 1. The relative area in the main and impurity phases determined within transmission integral approximation. The relative area of $\alpha$-Fe is below 1%. Fraction of iron atoms is proportional to the ratio of the relative area and the recoil-free fraction, see text.

| Crystalline phase<br>Absorber | FeTe$_{0.5}$Se$_{0.5}$ | Fe$_3$O$_4$ | Fe$_7$Se$_8$ |
|---|---|---|---|
| Fresh powder | 80(2) | 12(2) | 9(2) |
| After 19 days in air | 85(2) | 9(1) | 6(1) |
| After 4 months in air | 83(5) | 7(4) | 11(4) |
| After 6 months in air | 84(1) | 11(1) | 5(1) |
| After 16 months in air | 83.4(5) | 10.2(2) | 6.4(2) |
| Stored under argon, measurement at $T = 292$ K | 84.0(8) | 10.0(5) | 6.0(7) |
| Stored under argon, measurement at $T = 18$ K | 83.1(7) | 8.4(5) | 8.5(5) |

In the same way the data presented in figure 1 were treated. Surprisingly, within the period of 16 months of exposition of some absorbers to air, we could not observe, within the experimental accuracy, any systematic increase of the amount of magnetite, see table 1. The relatively lower statistics of the data shown in figure 1 (a) made the observation of impurity phases not possible. Although fresh powder was not measured with statistics as high as presented in the spectrum shown in figure 6 (a) for the sample not exposed to air and stored under argon atmosphere, the latter shows presence of impurity phases, thus indicating that they were present also in the original bulk sample.

In order to observe temperature dependence of recoilless fraction, the measurements in wide velocity range were performed together with an iron foil, both absorbers kept in the cryostat. The Fe foil of known thickness served as precise calibration device for setting the intensity scale [30]. Ratio of the area under the doublet to the area under the Fe sextet can easily be inferred from figure 7.

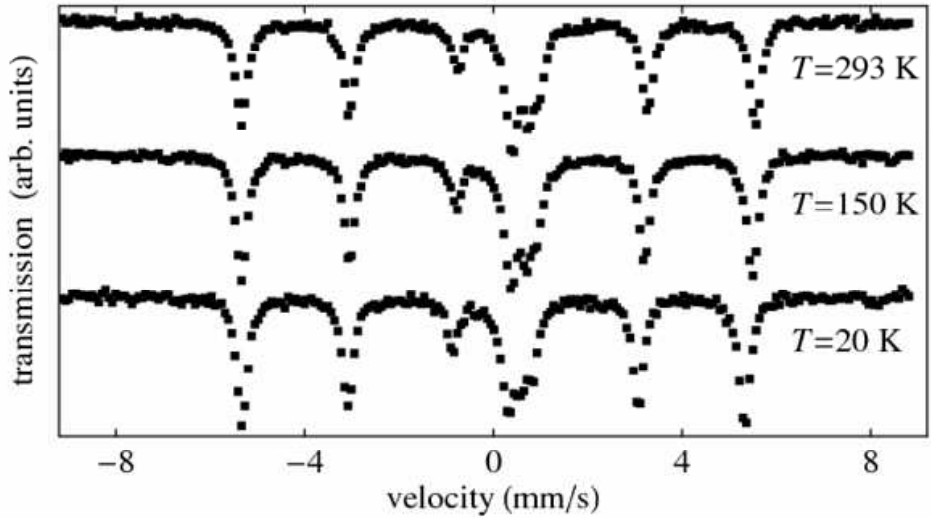

Figure 7. The Mössbauer spectra of $FeTe_{0.5}Se_{0.5}$ with planar density of the mass equal to 23(1) mg·cm$^{-2}$ and iron foil, both absorbers in the cryostat, measured for determination of the recoilless fraction.

## 4. Discussion

In the period of 16 months exposition of powdered absorber to ambient air conditions, we could not detect intuitively expected increase of the magnetite content, see figure 1 and table 1. The impurity contribution is hidden in the background for spectra measured with low statistics (figure 1(d)). It is thus clear that in order to extract the impurity phases from the main superconducting phase one has to perform the Mössbauer measurements with high statistics and cover wide enough velocity range as presented in figure 6. The overlapped peaks of the impurity phases $Fe_3O_4$ and $Fe_7Se_8$ form absorption shape similar to flat background, see figures 6 (c) and (d), which can be determined in measurements spanning wide velocity range. On the other hand, wide velocity range does not allow for precise determination of the doublet parameters. Thus both measurements, within wide and narrow velocity ranges, are required.

One can also expect, after evidence of easy formation of magnetite admixture [3], that grinding at ambient condition would cause easy oxidation to magnetite. Magnetization loops shown in figure 2 do not support such a simple mechanism. Indeed, the grinding causes a formation of some structural defects, which apparently changes magnetization processes seen in the case of a single crystal. Surprisingly, grinding under inert gas increases saturation magnetization whereas the grinding at contact with air does not increase the value of saturation magnetization. The results presented in table 1 show that in powder of a micrometer size, obtained from single crystal, a few percent of iron forms magnetite phase and amount of this phase remains unchanged in 16 months. Thus, one can exclude formation of magnetite as the result of chemical reaction of tetragonal $FeTe_{0.5}Se_{0.5}$ phase with atmospheric oxygen.

Figure 2 presents field dependence of magnetization for $FeTe_{0.5}Se_{0.5}$ single crystal and powder obtained by grinding under argon and ambient conditions recorded at 20 K in magnetic field up to 6 kOe. The value of saturation magnetization, $M_s$, for the studied monocrystal is of about 1 emu/g. Saturation magnetization of $Fe_7Se_8$, determined by Kamimura [31] at 80 K is of about 85 emu/cm$^3$, what corresponds to about 14 emu/g. Taking into account that the difference between saturation magnetization at 80 K and at 20 K is insignificant, we can estimate that the maximum volume fraction of $Fe_7Se_8$ phase in the studied sample does not exceed 7%. Its correlates well with 5.35(40)% estimation of the volume fraction for impurity hexagonal $Fe_7Se_8$ phase (space group $P6_3/mmc$) in $FeTe_{0.5}Se_{0.5}$ obtained from neutron powder diffraction measurements performed on similar crystal [32]. Saturation magnetization of $Fe_3O_4$ nanoparticles depends on their size and decreases with decreasing nanoparticle size. For the smallest nanoparticles with the size of about 5 nm, magnetization at 20 K in magnetic field of 6 kOe is of about 40 emu/g [33]. It means that volume fraction of the impurity $Fe_3O_4$ phase in the studied sample can not extend 2.5%. Powder X-ray data for the studied samples do not show peaks neither for $Fe_7Se_8$ nor $Fe_3O_4$, what gives upper estimation of their volume fraction to 5%.

Mechanically formed absorbers from powder of $FeTe_{0.5}Se_{0.5}$ show remarkable crystalline texture, similar in all absorbers prepared in this project, see examples in figure 4(d). This is observed in our powder X-ray measurements as enhanced intensities of (00$l$) Bragg peaks, reported earlier by Bendele *et al.* [32] in neutron powder diffraction experiment.

$Fe_3O_4$ and $Fe_7Se_8$ phases were detected in the investigated samples. The relative amount of these compounds was determined by fitting a function which is the sum of the subspectra with shapes as given in publications [24, 29]. The area under $Fe_7Se_8$ subspectrum strongly increases with decreasing temperature. This is an indication that $Fe_7Se_8$ forms small superparamagnetic particles. Indeed, at low temperature all $Fe_7Se_8$ particles show magnetic splitting seen as the subspectrum in figure 6(d). At room temperature a fraction of $Fe_7Se_8$ particles is thermally activated and does not show magnetic splitting present in bulk $Fe_7Se_8$ sites. Thermally activated fraction contributes to the main doublet in the Mössbauer spectrum. Because the isomer shift and the quadrupole splitting of $Fe_7Se_8$ sites are different from the ones of $FeTe_{0.5}Se_{0.5}$, the shape of the main doublet is thus slightly distorted. Our results are also in agreement with neutron measurements performed on single crystal of $FeTe_{0.5}Se_{0.5}$, where $Fe_7Se_8$ was detected [32], as well as with the Mössbauer and X-ray measurements performed on polycrystalline sample of $FeTe_{0.5}Se_{0.5}$, where $Fe_7Se_8$ was detected too [7].

The shape of the main $FeTe_{0.5}Se_{0.5}$ component agrees with the one reported for FeSe [3, 34, 35]. We do not observe magnetic order at Fe sites. Small abnormalities are observed at various temperatures. They are observed for the quadrupole splitting (figure 4 (a) at temperatures of about 70 K and 100 K), in a jump of the full width at half maximum at temperatures of about 50 K (figure 4(c)), and maximum in asymmetry of the main doublet at temperatures of about 130 K (figure 4(d)). These abnormalities were not detected in the first experiments performed with lower statistics (open points in figure 4). There are two possible explanations of these observations. The first one consists in a possibility of tetragonal to orthorhombic structural phase transition reported in [36] for FeSe at $T$ of about 70 K or at $T$ of about 100 K [37], and for $Fe_{1+\delta}Te_{0.43}Se_{0.57}$ at $T$ of about 40 K [38]. Orthorhombic distortion can naturally lead to a change in the electric field gradient acting on the $^{57}$Fe nucleus, resulting in changes of the quadrupole splitting. However, because observed changes are small, one cannot exclude another possibility that the detected unambiguously impurity phases of $Fe_3O_4$ and $Fe_7Se_8$ are responsible for observed non monotonous variation of hyperfine parameters. $Fe_3O_4$ exhibits the Verwey transition at about $T = 120$ K while $Fe_7Se_8$ undergoes spin reorientation at $T = 125$ K. At the phase transition points a change of the shape of the absorption spectrum occurs, influencing slightly the shape of the asymmetric doublet fitted to the spectra. Also, as already explained, the presence of a fraction of the superparamagnetic particles changes the shape of the main doublet.

A temperature change of the isomer shift at $T = 105$ K in superconducting FeSe was reported in [35]. Within the experimental precision we could not detect any peculiarities at this temperature in measurements on fresh powders as well as in measurements performed with higher statistics on aged samples (figure 4(b)). Instead, the systematic difference is observed between isomer shift of fresh and aged in argon samples, and its origin remains unclear

The sign of the EFG is found to be negative in agreement with [35]. Fitting $(1-aT^{3/2})$ temperature dependence to quadrupole splitting shown in figure 4 (a) one arrives at $a = 3.3 \cdot 10^{-5}$ K$^{-3/2}$ falling in typical range predicted by molecular dynamics calculations [39].

The Lamb-Mössbauer fraction does not show any peculiarities within the experimental accuracy. However, neither Debye nor Einstein model predictions [40] fit to the data shown in figure 8. This type of behaviour is observed in some systems and can be parameterized by temperature dependence of the Debye temperature [41, 42]. Our results correspond to recently reported anomalies observed in $FeTe_{0.5}Se_{0.5}$, where the Debye temperature determined from temperature dependence of the Lamb-Mössbauer factor was lower by about 200 K than those determined from the second order Doppler shift [7]. We also note that clearly nonharmonic behaviour of Fe dynamics was shown in layered LiFeAs superconductor [17].

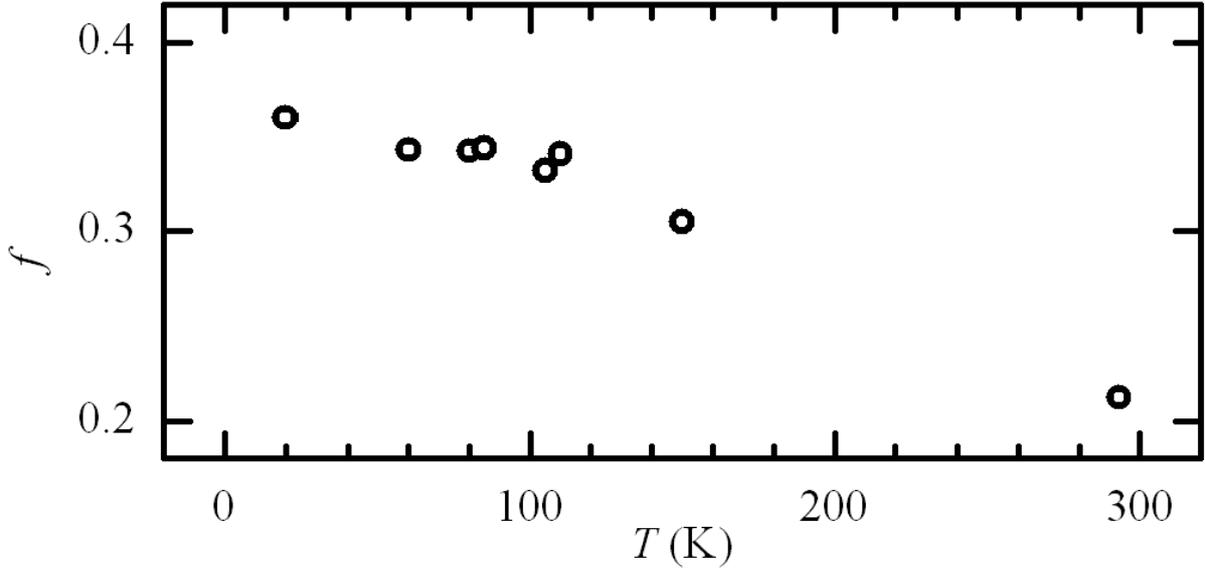

Figure 8. Temperature dependence of the Lamb-Mössbauer factor determined by internal standard method. Uncertainty is determined by precision of planar density of mass which is of about 4%.

Absolute values of recoilless fraction allow a quantitative estimation of the magnetite impurity phase. At room temperature 12(2)% of the area correspond to magnetite while 80(2)% to tetragonal $FeTe_{0.5}Se_{0.5}$ phase (first row data, table 1). Relative area is proportional to the concentration of $^{57}Fe$ per unit area and to the recoilless fraction. Taking recoilless fractions of magnetite $f_m = 0.89(4)$ [43], and of the tetragonal phase $f_{tetr}$ from figure 8 ($f$ at $T = 295$ K), the ratio of the amount of iron atoms in magnetite to the amount of iron atoms in tetragonal phase is $(f_{tetr}/f_m)·(12(2)\%)/(80(2)\%) = 0.035(6)$.

Another estimation, in which previously estimated recoilless fraction is not explicitly used, comes from figure 1. Magnetite sample with known thickness was used in this measurements. Analysis of spectra in figure 1 (f) results in area of magnetite equal to 52% whereas 42% of the total area is due to tetragonal phase. These values correspond to amount of tetragonal phase 21(1) mg·cm$^{-2}$ and extra magnetite 3.6 mg·cm$^{-2}$ (small amount of magnetite present in original sample is neglected). The ratio of masses is proportional to the ratio of areas, $3.6/21(1) = x·52/42$, where $x$ is some constant. In figure 1(a) area of magnetite is equal to 12(2)% and area of tetragonal phase 80(2)%. Similarly, one can write another proportionality: $s/21(1) = 12(2)/80(2)x$ with $s$ being planar density of mass of the magnetite in the sample corresponding to figure 1(a). One thus gets $s = (3.6$ mg·cm$^{-2})·(12(2)/80(2))·(42/52) = 0.44$ mg·cm$^{-2}$. Thus, for 0.44 mg·cm$^{-2}$ of $Fe_3O_4$ and 21(1) mg·cm$^{-2}$ of $FeTe_{0.5}Se_{0.5}$, one finds the ratio of Fe atoms in these two phases equal to 0.043(7). The two estimations are consistent within the experimental uncertainty and show also that low value of the recoilless fraction presented in figure 8 enhances relative area of a magnetite in the Mössbauer spectra. The Mössbauer estimations of the magnetite content are also consistent with results obtained from magnetization and diffraction measurements.

## 5. Conclusions

Experiments performed on powdered single crystals of $FeTe_{0.5}Se_{0.5}$ within broad velocity range revealed presence of minor content of two impurity phases: $Fe_3O_4$ and $Fe_7Se_8$. We have shown that quantitative estimation of the magnetite impurity phase is consistent with anomalously small value of recoilless fraction of tetragonal $FeTe_{0.5}Se_{0.5}$. The content of the $Fe_7Se_8$ phase obtained from fits with hyperfine parameters related to bulk crystals, depends strongly on temperature, indicating for its superparamagnetic behaviour. An increase of the magnetite content is not observed within period of 16 months exposition of the sample to air condition. Mechanical grinding apparently influences shape of the magnetization loops. Grinding under ambient air does not increase saturation magnetization in contrast to grinding under inert gas. The asymmetry of the absorption lines in the Mössbauer spectra depends on the sample orientation with respect to the wave vector of radiation, indicating for strong crystalline texture induced easily during absorber preparation. The samples stored in Ar atmosphere show small increase of the isomer shift of the doublet. Small peculiarities observed for temperature dependence of hyperfine parameters can be explained as due to either possible orthorhombic distortion similar to those observed in FeSe [36] or to the temperature behaviour of impurity phases $Fe_3O_4$ and $Fe_7Se_8$.

**Acknowledgements**
The work was partially supported by funds allocated for scientific research for the years 2008–2011 as a research project NN202172335 and by the EC through the FunDMS Advanced Grant of the European Research Council (FP7 'Ideas'). We thank Tomasz Dietl for suggesting this research and valuable discussions.